\documentstyle[11pt,paspconf]{article}
\def\kms{\ifmmode{$~km\thinspace s$^{-1}}\else km\thinspace s$^{-1}$\fi}
\begin{document}

\title{Faint star counts with HST}

\author{Chris Flynn}
\affil{Tuorla Observatory, Piikki\"o, Finland}
\author{Andrew Gould\altaffilmark{1}}
\affil{Department of Physics and Astronomy, Ohio State University, 
Columbus, USA}
\author{John Bahcall}
\affil{Institute for Advanced Study, Princeton, USA}

\altaffiltext{1}{Alfred P.~Sloan Foundation Fellow.}

\begin{abstract}
  
  We describe a program of star counts in the range $ 19 \la I \la 26$ made
with the WFPC cameras aboard the {\it Hubble Space Telescope}. Red ($V-I >
1.0$) stars at these magnitudes are primarily disk and spheroid M dwarfs. The
stars are found both on dedicated images as part of the parallel program and by
using appropriate archive data. We measure the faint end of the luminosity
functions of the disk and spheroid (i.e.~stellar halo). We measure the low mass
end of the mass function and show that M dwarfs do not dominate the total disk
or spheroid mass.  We place strong $I$ band constraints on the amount of halo
dark matter in the form of low mass stars (such as M dwarfs or cool white
dwarfs).  The disk and spheroid contribute only a minor amount of optical depth
toward the Magellanic clouds.

\end{abstract}


\section{Introduction}

  The {\it Hubble Space Telescope} (HST) has revitalised the classical field of
star counting, since it allows us to separate stars from galaxies more than
four magnitudes deeper than is possible from the ground. We have used HST to
search for faint red stars. There are four topics we address in searching for
such stars. 1) how much mass do M stars contribute to the disk and spheroid? 2)
does the Mass Function of the disk or spheroid rise at the faintest measurable
magnitude, possibly indicating large numbers of brown dwarfs beyond the
Hydrogen burning limit? 3) is the putative Galactic dark halo composed of faint
M dwarfs or cool white dwarfs? 4) how much do faint disk and spheroid M dwarfs
contribute to the microlensing signal seen toward the Magallenic clouds?

\section{Observations}

  Our stars our found on WFPC/WFPC2 images taken with HST in the F606W and
F814W filters. Part of the data is obtained under the Guaranteed Time Observer
(GTO) ``low latitude parallel program'' for which Ed Groth is Principal
Investigator.  The remainder is obtained by searching the Hubble Archived
Exposures Catalog (AEC{\footnote{http://stdatu.stsci.edu/aec.html}}) for fields
for which there are at least 2 exposures in both filters F606W and F814W. At
present (Oct 1998) our database consists of 17 fields obtained under the
parallel program, the Hubble Deep Field (HDF), the ``Groth Strip'' (28
contiguous fields) and 106 randomly collected fields from the Archive. Our data
base is increasing by a field every two or three weeks.  The distribution of
the fields in Galactic coordinates is shown in figure 1.  We only use fields
above 18 degrees Galactic latitude to avoid the uncertainties of reddening
corrections at lower latitude. The fields cover 4.4 square arc-minutes each at
a scale of 0.1 arcsec/pixel. Fields near globular clusters, clusters of
galaxies, dwarf satellites of the Milky Way and the LMC/SMC are not used.

\begin{figure}[!htb]
\plotfiddle{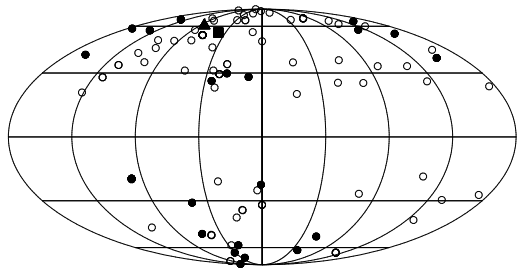}{45truemm}{0}{45}{45}{-150}{-70}
\caption{Distribution of the fields in Galactic coordinates.  The triangle
marks the Groth Strip; the square marks the Hubble Deep Field; filled symbols
--- fields analysed to date; unfilled circles --- fields still to be analysed.}
\end{figure}

  The images are stacked and cleaned of the large number of cosmic rays, using
a technique which treats each pixel individually, since HST's pointing and
tracking accuracy is very precise (Bahcall et al 1994). Although two images in
a filter are sufficient to remove most cosmics, three images or more is
best. In a small number of cases cosmics hitting the same pixel on both images
can conspire to produce a stellar like object in the cleaned image, but such
cases are removed by comparison of the cleaned and the original images.  A
typical star ($I = 21.8$, $V-I = 2.4$) is shown in figure 2. At high Galactic
latitude ($|b|>18^\circ$) there are only a few stars per field, so they must be
separated from the very numerous galaxies carefully. The canonical stellar
profile was found by determining the mean radial profile of over 30 stars found
on a very low latitude ($b=5^\circ$) image which is dominated by stars.  Tests
showed that the stellar profile is not dependent on location on the chips, so a
mean profile was adopted for comparison.

\begin{figure}[!htb]
\plotfiddle{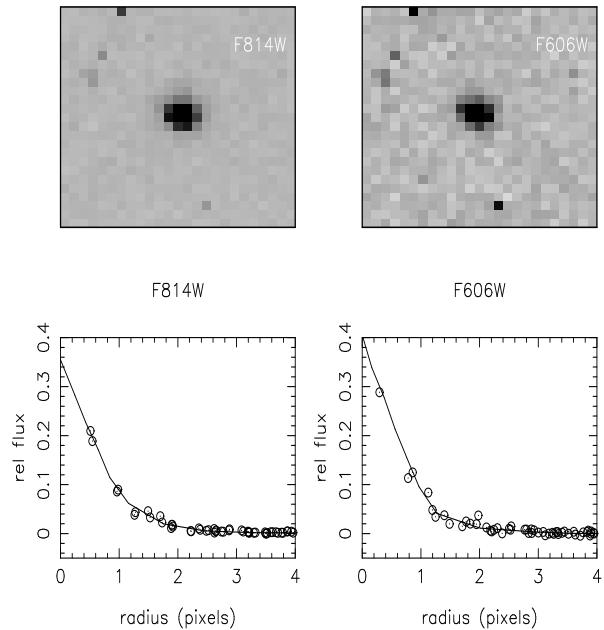}{85truemm}{0}{45}{55}{-130}{-30}
\caption{A typical star as seen on HST/WFPC. The radial profiles in F814W and
F606W are shown in the lower panels, and the images in the upper panels.  The
star is a close match to the mean stellar profile of many stars taken from a
low latitude image (shown as a solid line).}
\end{figure}

  A typical compact galaxy ($I=23.3$) is seen in figure 3. This is a
particularly compact example; most of the galaxies are considerably larger. A
small number of almost unresolved sources turned up, in particular in the
HDF. These were invariably blue ($V-I < 0.3$), and so had no impact on our
search for red stars.  However, their nature remains unclear: they may be small
(less than 1 kpc sized) star-burst regions (Elson, Santiago and Gilmore 1997).

\begin{figure}[!htb]
\plotfiddle{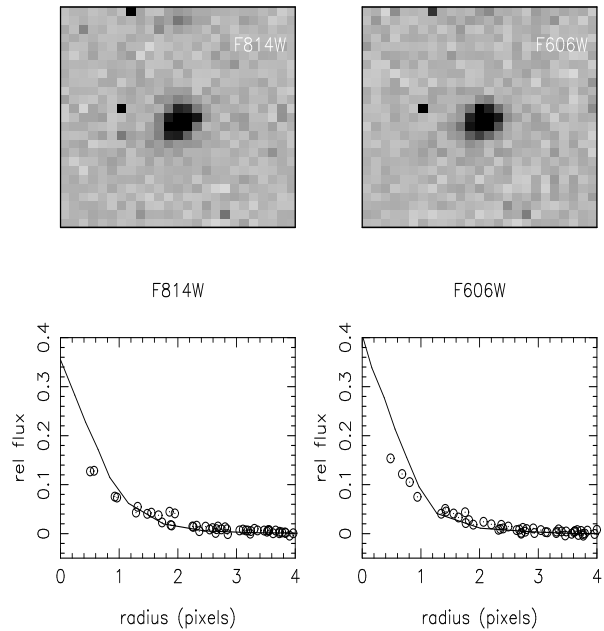}{85truemm}{0}{45}{55}{-130}{-30}
\caption{Typical compact Galaxy seen on HST. The radial profile of the 
galaxy falls well under the stellar profile (shown as a solid line).}
\end{figure}

  Aperture photometry was performed on the stars, and F606W and F814W
magnitudes were transformed to $V$ and $I$ magnitudes. The transformations to
$V$ and $I$ were determined by convolving the filter responses of F606W, F814W,
$V$, $I$ and individual chip response curves to stars in the library of Gunn
and Stryker (1983), since no empirical calibrations were available.  A similar
technique due to Holtzman et al (1995, 1998) yielded similar, although not
identical transformations. A source of concern was that the reddest stars in
the library were giants and not dwarfs.  Fortunately, an empirical
transformation has become available very recently (Salim and Gould 1998), who
obtained ground based $V$ and $I$ colours of red stars in the Groth strip. Both
the Bahcall et al and the Holtzman et al calibrations were found to need
systematic corrections at the 0.1 magnitude level, depending somewhat on
colour.  This impacts in a minor way our early results for the luminosity and
mass functions of M stars.

  Binaries are essentially unresolved in this program, since the stars are
typically 2 kpc distant. Binaries would be detected only for separations of
greater than 600 pc, at most a few percent of systems.

  We selected stars for which the signal to noise (S/N)$ > 12,$ based on counts
in an aperture of radius 4 pixels. At S/N $\approx 8$, sources clearly become
too noisy to separate into stars and galaxies. For fitting the mass and
luminosities functions we further restricted ourselves to stars 0.3 magnitudes
brighter than the magnitude limit set for each field by the S/N = 12.0
criterion. Our stellar/galaxian separation has been tested externally, at least
for the brighter sources, by Keck spectroscopy of 100 sources in the HDF. We
predicted the 4 stars and 96 galaxies in this sample correctly, as part of the
Caltech Redshift Challenge (Hogg et al 1998). The S/N of these ``bright'' ($I<
22.3$) stars was very high, however, well above the HDF magnitude limit at
$I=26.3$.
 
\section{I band limits on baryonic dark matter candidates}

  The Hubble Deep Field (HDF) was a superbly useful field for star counting.
Three groups found the same 18 stars in the HDF (Flynn, Gould and Bahcall,
1996; Elson, Santiago and Gilmore 1996; Mendez et al 1996).  Flynn, Gould and
Bahcall (1996) analysed the colour distribution of these 18 stars -- no stars
were found redder than $V-I = 1.80$ within 1.67 magnitudes of the magnitude
limit at $I=26.3$. This places strong limits on the $I$ band luminosity of
stellar objects which could form all or part of the Galactic dark halo.  In
particular, less than 1\% of the dark halo could be composed of M dwarfs if
they have the same $I$ band luminosity as disk M dwarfs of the same
colour/mass. Even stronger limits (i.e.~much less than 1\%) can be placed if
the dwarfs are assumed to have $I$ band luminosities typical of theoretical
models of metal poor stars (Graff and Freese 1996).  Although less attractive
than M dwarfs, another candidate for the halo dark matter is old, cool white
dwarfs, in particular because the microlensing searches indicate a typical lens
mass of about 0.5 $M_\odot$. Such old white dwarfs might lie along a linear
extrapolation of the sequence of old disk white dwarfs in the $M_I$ versus
$V-I$ CMD (see e.g.~Monet et al 1992) If so, then they would have to be at
least two magnitudes fainter than the faintest ($M_V \approx 16$) disk white
dwarfs in order not to be seen in large numbers in the HDF. Such white dwarfs
may not be red at all in which case they could actually be seen but
unrecognised in HDF (Hansen 1998).  More generally, Flynn, Gould and Bahcall
(1996) develop an $I$ band limit to the amount of baryonic matter in the form
of red ($1.8 < V-I < 4.8$) stars -- see their Eqn 4.2.

  HDF star counts can also be used to constrain the density of ``intracluster''
stars in the Local Group, i.e. outside the galaxies.  Intracluster stars of
this type have recently been identified in the Virgo cluster as Planetary
Nebulae (Feldmeier, Ciardullo \& Jacoby 1998).  Local Group giants and
subgiants $M_I \la 1.5$ and $0.6 \la V - I \la 1.5$ could be seen in HDF to a
distance $d \approx 0.9$ Mpc. In HDF there are no more than 2 such stars in the
the outer 90\% of the observed volume implying that their number density must
be less than $7 \times 10^{-11}$ pc$^{-3}$ at the 95\% confidence level.  This
constrains the ratio of the densities of Local Group to Galactic spheroid stars
to be less than 1/3000, about an order of magnitude lower than the limits
obtained by Richstone et al. (1992).

\section{Disk Luminosity Function and Mass Function}

  A useful field for illustrating our techniques in measuring the faint end of
the luminosity function is the ``Groth Strip'', a sequence of 28 contiguous
fields at $l=96$, $b=60$ (Rhodes et al 1997), for which star/galaxy separation
was possible in the range $18.75 < I < 23.79$.  These fields were primarily
taken for studying galaxy clustering, but their stellar content is particularly
useful because we obtain a large number of red stars at a single position on
the sky. In figure 4 we show the absolute magnitude versus vertical distance
modulus of the stars, assuming that each star lies on the empirically well
calibrated disk main sequence for M dwarfs of Monet et al (1992), $M_V = 2.89
+3.37(V-I)$.

\begin{figure}[!htb]
\plotfiddle{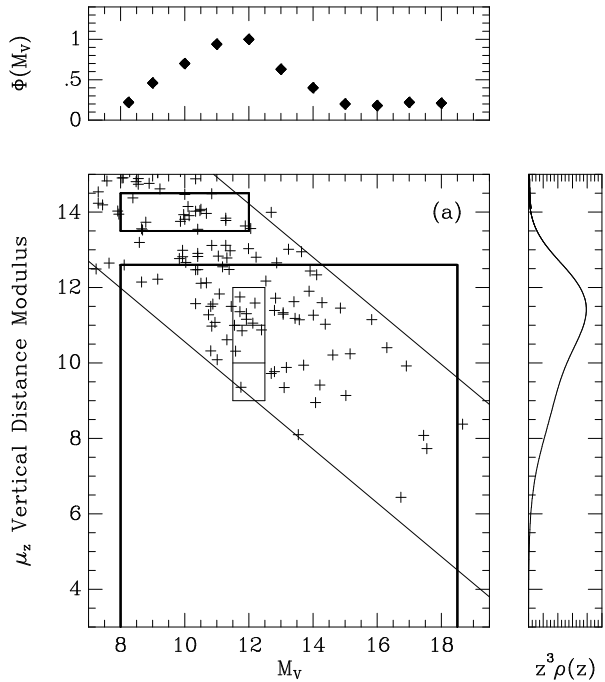}{95truemm}{0}{65}{65}{-150}{-30}
\caption{(a) M dwarfs in the Groth strip. Crosses mark the detected stars
plotted as absolute magnitude $M_V$ versus vertical distance modulus above the
plane ($\mu_z = V - M_V +5\thinspace{\mathrm \thinspace log \thinspace sin}
\thinspace b$). The sloping lines mark the bright and faint apparent magnitude
limits of the field. Stars in the large box marked by a heavy line are disk M
dwarfs, while spheroid M dwarfs lie in the smaller heavy outlined box. The
number of stars in the small 1 mag by 1 mag boxes is the product of the
vertical density function (right panel) and the LF (upper panel).}
\end{figure}

  These data are unique in constraining the total disk column of M dwarfs
rigorously, since HST's star/galaxy separation allows us to see all the M
dwarfs along any line-of-sight well out into the Galactic spheroid, beyond the
``top'' of the disk, as a result of which Malmquist bias corrections are
remarkably small. The disadvantage is that binaries are not resolved, so that
the binary correction to the observed LF must be made on the basis of binary
fractions in nearby M dwarfs. The results of an analysis of the Groth strip and
25 other fields (Gould et al 1997) show that the observed LF drops rapidly
after peaking at $M_V \approx 12$. Vertical density distributions were fit to
the M dwarf counts within 1 kpc, with a best description being a combination of
a shorter scale height sech$^2(z/z_0)$ ``thin disk'' and larger scale height
exp$(z/z_1)$ ``intermediate component'', where $z_0 = 320 \pm50$ pc and $z_1 =
640 \pm 60$ pc with relative fractions locally of 0.78 and 0.22
respectively. We note that these scale heights have no particular physical
significance, since their relative normalisations and the scale heights are
strongly anti-correlated. The important point is that the derived LF is
insensitive to the model of the vertical distribution of the stars.

  The observed LF drops rapidly below $M_V = 12$. We compare our LF with other
(ground based) measurements of the LF in figure 5. These are all measured
locally, whereas we see M dwarfs at distances of typically 2 kpc.  Our LF is
marginally inconsistent with the ground based LFs of Wielen et al (1983) and
Reid et al (1995). Part of the difference is certainly due to a correction to
the faint part of our LF for missing binaries.  M dwarfs in the disk have about
10\% binarity at the peak of the LF and about 50 \% binarity at the faint end
(Reid et al 1995, Gould et al 1995).  Correcting with these numbers raises the
falling part of the LF, although not by enough to make it flat at the faint end
(Gould et al 1997).

\begin{figure}[!htb]
\plotfiddle{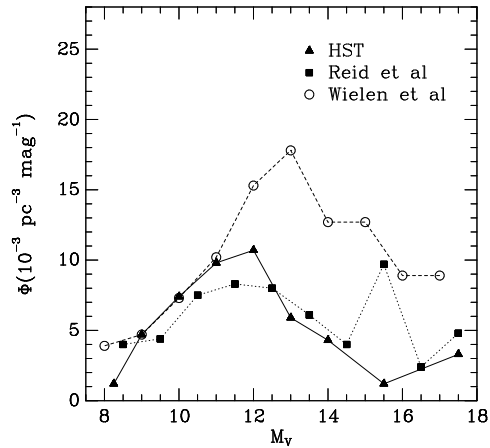}{60truemm}{0}{40}{40}{-130}{-95}
\caption{Disk luminosity function from HST star counts (binary uncorrected)
shown as triangles, compared to the ground based determinations by Wielen et al
(1983, circles) and (binary uncorrected) Reid et al (1995, squares).}
\end{figure}

  The LF can be converted into a mass function. In the case of disk M dwarfs,
this is relatively straight forward because good empirical mass-luminosity
relations exist (Henry and McCarthy 1993). We find a falling MF in the range $
M < 0.6\, M_\odot$ (Gould, Bahcall and Flynn 1997, figure 3). Although the
binary correction at the faint end of the MF may be enough to raise the falling
section so that it becomes flat, the evidence appears to be against the MF
rising into the regime below the Hydrogen burning limit, suggesting that there
are not a large number of brown dwarfs still to be found in the local
disk. This is consistent with recent studies which have finally turned up the
long sought free floating brown dwarfs in the nearby disk, although they are
unlikely to dominate the disk mass (Fuchs, Jahrei{\ss} and Flynn 1998). We
estimate from our MF the total contribution of M dwarfs to the disk mass as 12
$M_\odot $pc$^{-2}$: M dwarfs contribute only modestly to the total observed
disk column density of 40 $M_\odot $pc$^{-2}$, of which about 13 $M_\odot
$pc$^{-2}$ is in gas and 15 $M_\odot $pc$^{-2}$ is in non-M dwarf stars (Gould,
Bahcall and Flynn 1996). As a result of the low number of M dwarfs a marginal
discrepancy still remains between the observed (visible) mass and the
kinematically determined mass of the disk, estimates of which range from $46
\pm 9\, M_\odot pc^{-2}$ (Kuijken and Gilmore 1989) and $52 \pm 13\, M_\odot
$pc$^{-2}$ (Flynn and Fuchs 1994) to $ 84 \pm 25\, M_\odot $pc$^{-2}$ (Bahcall
et al 1992).

\section{Spheroid Luminosity Function and Mass Function}

  We have measured the LF of the spheroid (i.e.~stellar halo) from our most
distant dwarfs. While most of our M dwarfs are within a few kpc of the Galactic
disk, a significant number are at large distances, up to 30 kpc.  Distances to
all our dwarfs are assigned on the basis of the colour-luminosity relation for
(metal rich) disk M dwarfs. 

  We select a sample of spheroid M dwarfs by taking all objects which would be
higher than 8 kpc above the Galactic disk, if they were disk M dwarfs. At such
high $z$ heights we do not expect to see any disk stars at all (the scale
heights of the disk dwarfs were $320 \pm50$ pc and $640 \pm 60$ pc from the
previous section).  These must therefore be spheroid dwarfs, subluminous and
hence actually closer than they appear. We model the spheroid stellar density
distribution $\nu(x,y,z)$ as a power law

$$ \nu(x,y,z) = {({{x^2+y^2+(z/c)^2}\over{R_\odot}})^{-l/2}}$$

where $(x,y,z)$ are the Galactic coordinates, $R_\odot$ is the distance of the
sun from the Galactic center, $c$ is the spheroid flattening and $l$ is the
slope of the power law. 

  Since we do not know the metallicity and hence luminosity of any given M
dwarf, we assign density distributions to each object along the line-of-sight
based on an assumed metallicity distribution of the spheroid and adopted
metallicity dependent colour-magnitude relations (CMR). The CMR relies
primarily on the CMD of 43 high velocity M subdwarfs (tangential speeds greater
than 260 \kms) kindly made available to us in advance of publication by C.~Dahn
(itself an update of Dahn et al 1995). There are unfortunately not quite enough
of these stars to sample the CMD densely enough to empirically calibrate the
CMR, and we rely to some extent on the theoretical isochrones of Baraffe et al
1997 in order to interpolate within the CMR. Metallicities are not yet
available for the M subdwarfs in sufficient number to be completely sure that
there are not still systematic offsets between the data and the isochrones,
however for our purposes this is secondary because the isochrones are only used
to interpolate between the data points.

  The spheroid LF is recovered via maximum-likelihood techniques from the star
counts. There are 166 spheroid M dwarfs in 53 fields. Our best fitting Galactic
parameters are $c=0.82 \pm 0.13$ and $l = 3.13 \pm 0.23$, where we have set
$R_\odot = 8$ kpc.  Our LF is shown in figure 6 where it is compared with
spheroid LFs determined using a number of ground based techniques, by Dahn et
al (1995 --- DLHG) and by Bahcall and Casertano (BC --- 1986). A less than
desirable feature of the LF determination is that neighbouring bins have
correlation coefficients of about $-0.3$, which physically corresponds to the
fact that stars could almost equally well be attributed to the bin on either
side of their assigned luminosity bin.  This is a physical consequence of the
range of metallicity and hence absolute magnitude amongst spheroid stars, and
cannot be circumvented without individual metallicity or luminosity estimates
for the stars (e.g.~via multiband photometry).

\begin{figure}[!htb]
\plotfiddle{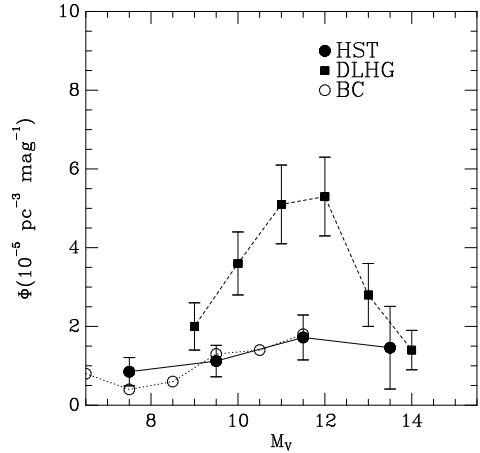}{60truemm}{0}{40}{40}{-130}{-85}
\caption{Spheroid LFs. Filled circles show the LF derived from HST star counts
for stars typically at distances of 5 to 10 kpc above the disk.  The other
symbols are for ground based LFs of the nearby spheroid.  Open circles :
Bahcall and Casertano (1986), filled squares Dahn et al (1995).}
\end{figure}

  We note briefly that the comparison LFs of DLHG and BC have both been
corrected by us (Gould et al 1998) for incompleteness factors arising as a
consequence of the kinematic (proper motion limited) selection of their
spheroid samples and consideration of an ``intermediate component'' (Casertano,
Ratnatunga and Bahcall 1990).


  The HST and ground based (DLHG) LFs appear different. Is the difference
significant?  Solving for the best fitting LF which minimises $\chi^2$ for both
samples, where the anti-correlation in the bins in the HST sample is carefully
taken into account, we find that the LFs differ at the $2.8\sigma$
level. Nevertheless, we hesitate to conclude that the LFs are different.  We
consider it likely that as samples of M subdwarfs with individually determined
abundances become available, systematic corrections to both the ground based
and HST LFs may emerge, and it would only take small systematic errors to
reduce the discrepancy between the LFs. The difference may be real, and a
possible explanation is that the locally determined spheroid LFs sample part of
a quite flattened spheroidal component to which we would not be sensitive at
large distances from the plane (Hartwick 1987, Sommer-Larsen and Zehn 1990).

  Having noted potential pitfalls in the determination of the spheroid LF, we
now boldly proceed to the spheroid mass function (MF). Converting the LF to an
MF for the spheroid is considerably more difficult than for the disk, because
we now must rely on theoretical calculations of the mass-luminosity relation
rather than the excellent empirical relation available for the disk. As a
consequence, any systematic errors in the theoretical isochrones will propagate
into the determination of the MF.

  We cannot simply convert the spheroid LF to an MF, because the observables
for our spheroid sample are colour and apparent magnitude rather than
luminosity, and the spheroid is composed of a wide range of metallicity and
hence mass at a given colour. Formally, we use the same maximum likelihood
technique as is used to recover the LF from the star counts, using colour-mass
relations rather than colour-luminosity relations.

 In brief, over the mass range $0.09 < M/M_\odot < 0.71$, we find no obvious
signs of structure in the mass function, and find that it can be characterised
by a power law, $dN/d ln M \propto M^\alpha$, with $\alpha = 0.25 \pm 0.32$.
This mass function is not corrected for binaries.  A steepening (decrease in
the power law exponent) of the low mass part ($M < 0.6\, M_\odot$) of the MF for
the disk of $\approx 0.35$ can be expected in the presence of binaries.  The
binary fraction in the spheroid is as yet unknown, but assuming that the binary
fraction is the same as in the disk, the spheroid MF would be approximately
flat ($\alpha = 0$) or slightly rising for $M < 0.6\, M_\odot$.

  Our measurement of the spheroid MF almost down to the Hydrogen burning limit
allows us to estimate the mass density of the spheroid. The major uncertainty
in such an estimate is the amount of material that is trapped in stellar
remnants (white dwarfs) and the amount of sub-stellar matter, neither of which
are yet well constrained by observations. We can extrapolate the power-law to
zero mass in order to estimate the amount of matter in low mass ($M < 0.71 $)
objects. Ignoring the binary correction, this yields a local spheroid matter
density of $\rho_{\rm sph} = (2.9 \pm 0.9)\times 10^{-5}\, M_\odot $pc$^{-3}$,
while adopting a correction for binaries based on the binary fraction in the
disk yields a mass density of $\rho_{\rm sph} = (3.6 \pm 0.9)\times 10^{-5}\,
M_\odot $pc$^{-3}$, a 25\% increase. Fuchs and Jahrei{\ss} (1998) have
determined a lower limit to the halo mass density by isolating spheroid stars
within 25 pc using Hipparcos parallaxes. After selecting stars in the mass
range $0.09 < M/M_\odot < 0.71$ from their Table 1 and using only high
tangential velocity stars ($V_T > 220$ \kms) with an appropriate correction
factor for incompleteness, their lower limit to the density is $(3 \pm 2)
\times 10^{-5}\, M_\odot $pc$^{-3}$, quite similar to our estimate.

\section{Contributions of disk and spheroid to the microlensing}

  The Microlensing discovered towards the Magallenic clouds in the last few
years is a major step forward in dark matter studies. A significant fraction of
the putative dark matter may be in the form of compact objects of order 0.5
$M_\odot$ (Alcock et al 1997). Our studies of the disk and spheroid MF allow us
to estimate the amount to which directly detected stars within these structures
contribute to the observed optical depth towards the LMC.

  For the disk, the total visible column density is 40 $M_\odot $pc$^{-2}$
(where 12 $M_\odot $pc$^{-2}$ is in M dwarfs). We characterise the density
distribution of the M dwarfs by a sum of a sech$^{2}$ and an exponential (see
section 4). The amount of optical depth along a line-of-sight at Galactic
latitude $b$ for stars with density distribution $\rho(z)$ is given by $\tau =
{\mathrm csc}^2 b \int_{0}^{\infty} 4\pi G \rho(z) z c^{-2} dz$ where $z$ is
the vertical height above the disk. We estimate an optical depth toward the LMC
$b=-33^\circ$ due to disk stars of $\tau_{\rm disk} = 8 \times 10^{-9}$.  The
latest measurement of the optical depth toward the LMC is $2 \times 10^{-7}$
(Alcock et al 1997); hence the disk contributes approximately 4\% to the
microlensing signal.

  For the spheroid, we derived in section 5 a local mass density in low mass
stars of $3.6 \times 10^{-5}\, M_\odot $pc$^{-3}$, whereas the local density of
the dark halo is of order $9 \times 10^{-3}\, M_\odot $pc$^{-3}$ (e.g.~Bahcall
1984). Hence the spheroid contributes less than 1\% of the microlensing signal.

\section{Future prospects}

  The luminosity and mass functions reported here for the disk and spheroid are
based on stars detected in about 40 fields. We currently have almost 100 more
fields which are under analysis, with additional fields being added at the rate
of about 2 dozen per year.  While the major trends are clear, these new data
will allow us to beat down the Poisson noise, particularly at the faint end of
the LF. Colours of the stars, previously transformed to $V$ and $I$ based on
stellar spectra and band passes, have now been calibrated from the ground ---
this will lead to small changes in our derived LFs and MFs. A few years have
elapsed since the Groth Strip images were taken: reimaging all or part of the
Groth Strip (or HDF) would allow us to measure proper motions for a large
number of faint M dwarfs, giving kinematic information and allowing a better
accounting of the fractions contributed by the disk and the spheroid. Near-IR
imaging of the Groth Strip would allow a measurement of metallicities of the M
dwarfs, which is possible using JHK photometry (Leggett 1992).  Very deep low
latitude images ($|b| < 30^\circ$) would better constrain the faint end of the
disk LF, as at present most of our very deep images are at high Galactic
latitude. Star counts with HST could also be used to constrain models of the
warp/flare of the outer disk, recently proposed as an alternative explanation
of the microlensing (Evans et al 1998).

\acknowledgments

  We have benefitted greatly from discussions with Neill Reid and Conard
Dahn. Isabelle Baraffe and Gilles Chabrier helped clarify several issues
related to theoretical isochrones. Conard Dahn and Isabelle Baraffe kindly made
available some of their work in advance of publication. Ed Groth kindly gave us
access to his HST parallel data in advance of publication.  Dante Minniti
suggested to us following up proper motions in the Groth Strip. AG was
supported by NSF grant AST 97-27520 and in part by NASA grant NAG 5-3111. JNB
was supported by NASA grant NAG 5-1618. CF thanks the IAS for a travel grant
and support during several stays in Princeton and Columbus.  The work is based
in large part on observations with the NASA/ESA Hubble Space Telescope,
obtained at the Space Telescope Science Institute, which is operated by the
Association of Universities for Research in Astronomy, Inc. (AURA), under NASA
contract NAS5-26555.  Important supplementary observations were made at KPNO
and CTIP operated by AURA.

\end{document}